\documentclass[aps,twocolumn,longbibliography] {revtex4-2}
\usepackage{amsmath,amssymb,graphicx,amsfonts}

\usepackage{times}
\usepackage[colorlinks=true,linkcolor=blue,urlcolor=blue,citecolor=blue,breaklinks=true]{hyperref}

\usepackage{comment}

\usepackage[utf8]{inputenc}

\def\da{\dagger}

\def\eq#1{Eq.~(\ref{#1})}
\def\da{\dagger}

\date\today
\begin{document}
	
	\title{Long-lived oscillating state in a Bose gas with an attractive polaron}
	
	\title{Long-lived oscillations of an attractive polaron in a Bose gas}
	
	\author{Saptarshi Majumdar}
	\author{Aleksandra Petkovi\'{c}}
	\affiliation{Univ Toulouse, CNRS, Laboratoire de Physique Th\'{e}orique, Toulouse, France}

	\begin{abstract}
		We study the out-of-equilibrium dynamics of an attractively interacting impurity suddenly immersed with a nonzero initial velocity into a system of one-dimensional weakly interacting homogeneous bosons. We uncover and characterize different dynamical regimes in the parameter space. Especially interesting is the relaxation of a fast impurity with a mass close to or exceeding the critical one, where the impurity exhibits undamped temporal long-lived velocity oscillations before reaching a stationary state.  The underlying mechanism is the transient localization of a boson depletion cloud near the impurity, that oscillates around the boson density peak situated at the impurity position. The lifetime of this oscillating state increases with the absolute value of the impurity-boson coupling. Cold atomic gases provide an ideal playground where this phenomenon can be probed.

	\end{abstract}

	\maketitle
	
\section{Introduction}

The motion of a distinguishable particle (an impurity) coupled to an environment is ubiquitous in physics. It describes many different fundamental phenomena \cite{landau+49,Devreese_2009,AnnalsKamenev}. A famous example is the notion of a polaron quasiparticle, introduced by Landau and Pekar \cite{landau+pekar},  who considered the motion of an electron in a dielectric crystal. The surrounding ions try to screen the electron charge, and severely modify the properties of the bare electron, which, together with a local phonon cloud, forms a polaron  \cite{Frohlich01071954}. The concept of polaron is nowadays broader and describes a mobile impurity dressed with a cloud of excitations of the host system. 
	
Understanding the out-of-equilibrium dynamics of a mobile impurity in a many-body environment  is a challenging problem. Particularly interesting are one-dimensional (1d) systems, where interactions can lead to very different physical phenomena with respect to their higher-dimensional counterparts \cite{Giamarchi}. In this work, we consider a 1d system of bosons. The interplay of the impurity-boson and the boson-boson interaction in out-of-equilibrium conditions leads to interesting dynamical phenomena \cite{zvonarev2007spin,gangardt2009bloch,QFlutterNature,knap2014quantum,robinson2016motion,PhysRevLett.117.113002,LewensteinBrownien,QuenchZvonarev,EnssPhysRevA.99.023601,10.21468/SciPostPhys.8.4.053, PhysRevLett.122.183001,EnssPhysRevResearch.2.032011,PhysRevLett.127.185302,atoms10010003,MyPRLdissipative,Will_2023,Quench,BlochOsc,Ion1d}. While the process of the Bose polaron formation in 1d geometry has been intensively investigated in the case of repulsive impurity-boson interaction \cite{zvonarev2007spin,QFlutterNature,knap2014quantum,LewensteinBrownien,PhysRevLett.122.183001,atoms10010003,Will_2023,Quench,YangGaudinFlutter}, this problem remains much less studied for an attractive impurity-boson interaction \cite{Mistakidis_2019,Ion1d}.
	
	In this work, we study the nonequilibrium dynamics of dressing and relaxation of an attractively interacting impurity suddenly immersed in the ground state of homogeneous bosons with a nonzero initial velocity. We take advantage of weak interaction between the bosons and solve the time-dependent mean-field equation of motion for bosons in the reference frame co-moving with the impurity.  
	We investigate the dynamical response of the bosons by monitoring the time evolution of their density and phase profile. It allows us to characterize the excitations emitted in the process of energy and momentum transfer from the impurity to the bath. Apart from the peak in the boson density at the impurity position, dispersive density shock waves and a depletion cloud also form, and in a general case, they move away from the impurity leading to a locally established  stationary state. We study the evolution of the impurity velocity in time, as a function of the impurity mass, the initial velocity and the impurity-boson interaction strength.  
	
The relaxation of a fast impurity with the mass close to or greater than the critical one is particularly interesting. After a rapid drop, the impurity velocity exhibits long-lived undamped temporal oscillations. This phenomenon is due to the trapping of the boson depletion cloud and its oscillations around the peak in the boson density situated at the impurity position. In this process, the depletion cloud gets progressively split into two parts positioned on the opposite sides of the impurity. The two depletion holes get repelled, and move away from the impurity, while the density peak continues the motion together with the impurity with a stationary velocity. The stronger the impurity-boson interaction is, the longer the lifetime of this oscillating state is. Cold atoms provide an excellent platform to probe the aforementioned phenomena \cite{2012quantum,Meinert945,JinPhysRevLett.117.055301,PhysRevLett.117.055302,HadzibabicPhysRevX.15.021070,grusdt2025impurities}.
	
The paper is organized as follows. We introduce the model in Sec.~\ref{sec:model}, and provide an analytic solution for a finite-momentum stationary ground state of the system, and its properties in Sec.~\ref{sec:Apolaron}. 
Section \ref{sec:relaxation} studies the relaxation dynamics of the system after the impurity injection into the bosonic bath, focusing on the influence of the initial impurity momentum, the impurity mass and the impurity-bath
coupling. In Sec.~\ref{sec:alternative}, we study the relaxation dynamics using an alternative equation of motion. We then report and characterize a special type of dynamics of fast and sufficiently heavy impurities in Sec.~\ref{sec:resonant}, where the impurity undergoes undamped long-lived oscillations before reaching a stationary state. The main conclusions are presented in Sec.~\ref{sec:disscusion}. Additional details on different stationary solutions are presented in Appendix \ref{sec:Appendix}.
		
	\section{Model\label{sec:model}}

	We study a system consisting of a single impurity immersed into a bath of 1d bosons at zero temperature. The system is modelled by the Hamiltonian
	\begin{align}
		\hat{H}=\frac{\hat{P}^2}{2M}+\hat{H}_b+G \hat{\Psi}^\da(\hat{X})\hat{\Psi}(\hat{X}).
		\label{eq:H}
	\end{align}
	The impurity has a mass $M$. Its momentum and position operators are denoted by $\hat{P}$ and $\hat X$, respectively. The Hamiltonian $\hat H_b$ describes interacting bosons and takes the form  
	\begin{align}
		\hat{H}_b=\int \mathrm{d}x\left[-\hat\Psi^\da(x)\dfrac{\hbar^2\partial_x^2}{2m}\hat\Psi(x)+\frac{g}{2}\hat\Psi^\da(x)\hat\Psi^\da(x)\hat\Psi(x)\hat\Psi(x)\right].
	\end{align}
	The single-particle bosonic operators  $\hat\Psi^\da(x)$ and $\hat\Psi(x)$ obey the commutation relation $[ \hat\Psi(x) , \hat\Psi^\da(x')]=\delta(x-x')$. The repulsion strength between the bosons is $g$. We introduce the dimensionless parameter $\gamma=m g/\hbar^2 n_0$, where $m$ is the mass of bosons and $n_0$ is their mean density. The last term in \eq{eq:H} models the impurity-boson interaction, where the impurity couples locally to the boson density with a coupling constant $G$.  We assume that the impurity-boson interaction is attractive, $G<0$.

	We perform the Lee-Low-Pines transformation \cite{LeeLowPines} $\hat{\mathcal{H}}=\hat{U}^\da \hat{H}\hat{U}$, where $\hat{U}=e^{-i \hat{X} \hat{p}_b/\hbar}$ and the momentum of the bosons is $\hat{p}_b=-i\hbar\int \mathrm{d}x  \hat\Psi^\da(x)\partial_x \hat\Psi(x)$. In the new referent system the impurity is situated at the origin and the Hamiltonian reads as
	\begin{align}\label{eq: HLee-Low}
		\hat{\mathcal{H}}=\hat{H}_b+\frac{(p-\hat{p}_b)^2}{2M}+G \hat{\Psi}^\da(0)\hat{\Psi}(0).
	\end{align}
	The total momentum of the system in the laboratory frame is denoted by $p$ and is conserved in time. At initial time, the impurity is free and has a moment $p=M V_0$, while the bosons are in their zero-momentum ground state. Then, the impurity-boson interaction is turned on, and we study the time-evolution from this far-from-equilibrium initial state.
	
In the following, we consider weakly-interacting bosons with $\gamma\ll 1$. In this limit, we employ the small-$\gamma$ expansion of the single-particle bosonic operator \cite{pitaevskii_bose-einstein_2003,sykes_drag_2009,CasimirNewJPhys}
\begin{align}\label{eq:expansion}
\hat\Psi(x,t)=\Psi_0(x,t)+\gamma^{1/4}\hat\Psi_1(x,t)+\ldots, 
\end{align}
where the field $\Psi_0(x,t)$ describes the condensate wave function in the absence of  fluctuations, while the higher order contributions account for the effects of quantum and thermal fluctuations.
In the leading order, one obtains the equation of motion for $\Psi_{0}$  \cite{PhysRevA.100.013619,atoms10010003,Will_2023}
\begin{align}
		i\hbar \partial_t{{\Psi_0}(x,t)}=\Bigg[&-\frac{\hbar^2}{2}\left( \frac{1}{m}+\frac{1}{M}\right)\partial_x^2+g |{\Psi_0}(x,t)|^2\notag\\&+G\delta(x)+i\hbar V(t)\partial_x\Bigg] {\Psi_0}(x,t).
		\label{eq:mean-field1}
\end{align}
Here, the impurity velocity $V$ takes the form 
\begin{align}
		V(t)=\frac{p}{M}+i\frac{\hbar}{M}\int \mathrm{d}x  \Psi_0^*(x,t)\partial_x \Psi_0(x,t).
		\label{eq:Vimp}
\end{align}

	\section{Attractive polaron\label{sec:Apolaron}}

	\begin{figure}
		\centering
		\includegraphics[width=0.9\columnwidth]{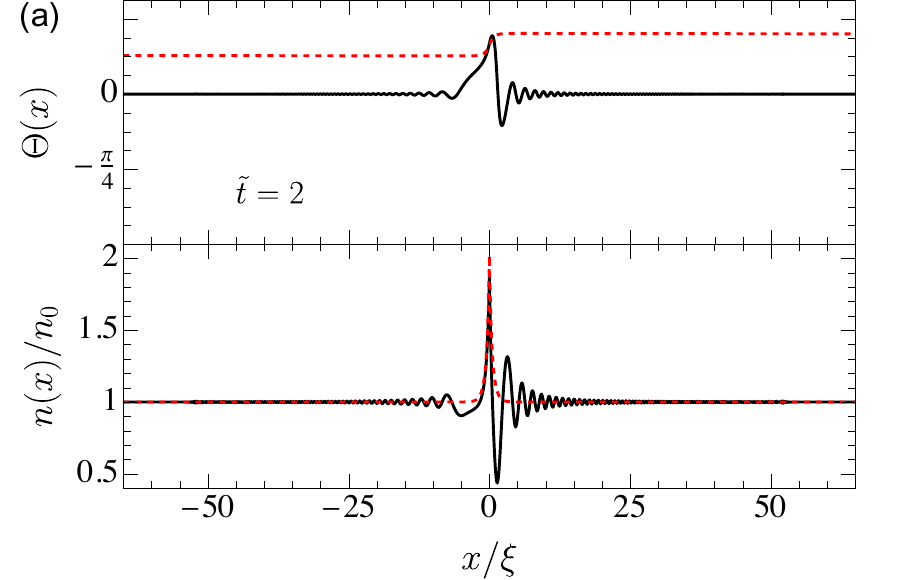}\\
		\includegraphics[width=0.9\columnwidth]{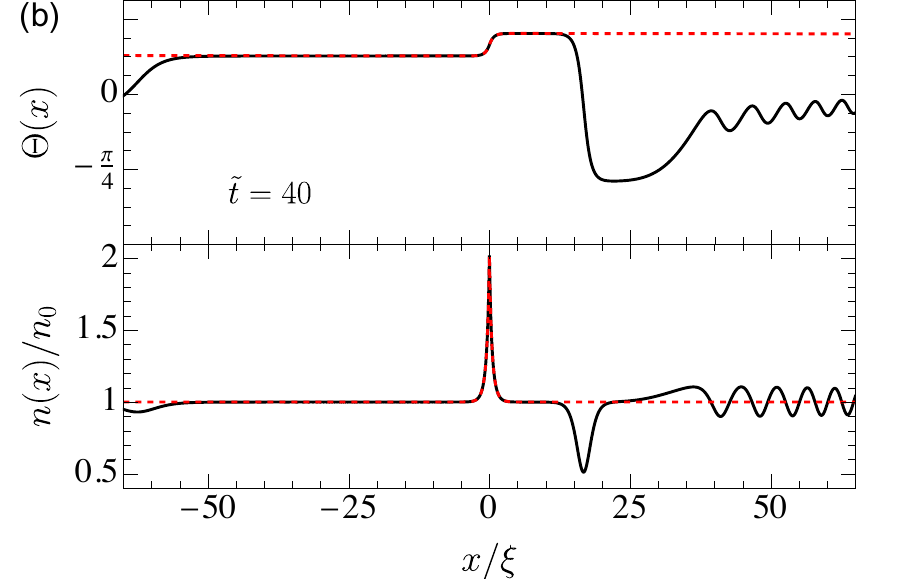}
		\caption{Time evolution of the phase $\Theta(x,t)$ and the density $n(x,t) = |\Psi_0(x,t)|^2$ of bosons in the vicinity of the impurity after a quench of the impurity-boson interaction, for $\tilde{G}=-0.8$, $M=3m$, and $V_0=1.5v$ at (a) $\tilde{t}=t g n_0/\hbar=2$ and (b) $\tilde{t}=40$. The red dashed lines denote the phase and the density of the ground state (\ref{eq:MFsolution}) for the numerically obtained final impurity velocity $V_f=0.43 v$. The corresponding impurity velocity evolution in time is shown in Fig.~\ref{fig4}(a). At both times shown, the impurity velocity is positive. Here, the system size is $L=3200\xi$. Thus, the contribution $\theta x/L$, that ensures the periodicity of the phase, appears as an almost straight line over the shown interval of $x$.}
		\label{fig1}
	\end{figure}
	
In this section, we consider a stationary solution of \eq{eq:mean-field1} subject to periodic boundary conditions. It gives the stationary state reached by the system in the protocol where the boson-impurity interaction is switched on adiabatically slowly.  The solution can be written as   \cite{pitaevskii_bose-einstein_2003}
	$
	\Psi_0(x,t)=\Psi_0(x)e^{-i \mu t/\hbar},
	$
	where $\mu$ denotes the chemical potential of bosons. In the stationary state, the impurity velocity is constant, $V(t)=V_f$. The solution can be evaluated analytically by a generalization of the result of Ref.~\cite{Hakim} to an attractive finite-mass impurity. It reads as 
\begin{align}
\Psi_0(x)=&\sqrt{\frac{\mu}{g}}\exp{\left[i\; \text{sgn}(x) \arctan\left(\frac{b \tanh{x_0}}{a}\right)+i\theta \frac{x}{L}\right]}\notag\\ &\times\left[a-i \;b \;\text{sgn}(x)\tanh{\left(\frac{b|x|}{\xi\sqrt{1+m/M}} +x_0\right)}\right]
\label{eq:MFsolution}
\end{align}
in the thermodynamic limit, for the total momentum of the system being in the interval $-\pi\hbar n_0<p\leq \pi \hbar n_0$.
Here the system length is $L$, the healing length is $\xi=\hbar/\sqrt{m \mu}$, and the sound velocity is $v=\sqrt{\mu/m}$. Additionally, $a={V_f}/{v\sqrt{1+m/M}}$ and $b=\sqrt{1-a^2}$.
The term containing $\delta(x)$ in \eq{eq:mean-field1} imposes the jump of the first spatial derivative of the wave function at the origin. From this constraint, it follows that the shift parameter $x_0$ satisfies
\begin{align}
		\frac{G}{\hbar v}\frac{1}{\sqrt{1+m/M}}(a^2+b^2 \tanh^2{x_0})=b^3 \tanh{x_0}\; \textrm{sech}^2{x_0}.
		\label{eq:xo}
\end{align}
Apart from the boson density, the impurity also modifies the phase of the condensate, causing a phase drop $\theta$ across its position. In the thermodynamic limit, $\theta$ reads as 
\begin{align}
		\theta=&2 \arctan \left(\frac{b}{a} \right) -2\arctan\left(\frac{b }{a}\tanh{x_0}\right).
		\label{eq:theta}
\end{align} 
Thus, the contribution $\theta x/L$ in the phase of the condensate wave function (\ref{eq:MFsolution}), takes care about periodic boundary conditions. Note that this term gives a contribution to the system momentum carried by the supercurrents, but does not contribute to the energy.

Equation (\ref{eq:xo}) admits solutions only for the impurity velocities smaller than the critical velocity, $|V_f|\leq v_c$. Contrary to the case of a repulsively interacting impurity \cite{Hakim}, the critical velocity does not depend on  $\tilde{G}=G/\hbar v$ here and reads as $v_c=v\sqrt{1+m/M}$. 
This result is in agreement with the findings for an infinitely heavy impurity \cite{PhysRevA.66.013610,PhysRevA.64.033602}.
Moreover, for $\tilde{G}<0$, there is just \emph{one} physical solution of Eq.~(\ref{eq:xo}). It leads to a higher boson density at the impurity position than the mean boson density, $n(0)>n_0$. This solution for $x_0$ is a complex number and satisfies $\tanh{x_0}>1$, with the imaginary part of $x_0$ being $i\pi/2$. Then,  $\tanh{\left(\beta |x| +x_0\right)}=\coth{\left[\beta |x| +\mathrm{Re}(x_0)\right]}$ where $\beta$ is a real number. Thus, the stationary solution (\ref{eq:MFsolution}) describes a peak in the boson density centred at the impurity position with $n(0)>n_0$. Contrary to the repulsive case, the phase drop $\theta$ is negative for $V_f>0$, implying the phase increase across the impurity position. The phase and the density of bosons are shown by the red dashed line in Fig.~\ref{fig1}. 

The other two solutions of Eq.~(\ref{eq:xo}) give $x_0\leq 0$. Thus, for a nonzero $x_0$, they give a peak in the boson density centered at $x=0$, but with $n(0)\leq n_0$,  and surrounded by a depletion region. Thus, they are not expected to describe the ground state of the system \cite{Parisi2017}. We provide a detailed analysis of these solutions in App.~\ref{sec:Appendix}.

	We consider a finite impurity coupling constant $\tilde{G}$. The chemical potential is calculated from the constraint
	$
	\int_{-L/2}^{L/2} \mathrm{d} x |\Psi_0(x)|^2=n_0 L,
	$
	and it takes the following form
	\begin{align}\label{eq:muEv}
		\mu=g n_0+2 b \hbar  \frac{1-\tanh{x_0}}{L} \sqrt{\frac{g n_0}{m}(1+m/M)} .
	\end{align}
	Here $b$ and $x_0$ are evaluated by replacing $\mu$ by its leading order term $\mu_0=gn_0$. We are not interested in finite size corrections. Thus, from now on, all the parameters will be evaluated using $\mu_0$ instead of $\mu$, if not stated differently. We have evaluated the first finite-size correction in $\mu$ because it is needed for the subsequent evaluation of the energy in the thermodynamic limit. 
	
	Using Eq.~(\ref{eq:Vimp}), we obtain the relation between the initial impurity momentum $p=M V_0$  and its final value  $MV_f$, that reads as 
	\begin{align}
		p=&M{V}_f - {2\hbar n_0}ab \left(1 - \tanh{x_0} \right) +\hbar n_0 \theta + 2 k\pi\hbar n_0,
		\label{eq:momentum}
	\end{align}
	for $\pi \hbar n_0 (2k-1) <p\leq\pi\hbar n_0(2 k+1)$. Here $k$ is an integer. Note that increasing $V_f$ from zero to $v_c$, the density peak at the impurity position increases, as well as the absolute value of the phase drop across the impurity. We emphasize that $V_f$ satisfying \eq{eq:momentum} is not the final impurity velocity for the post-quench dynamics, as will be shown in Sec.~\ref{sec:relaxation}. Equation (\ref{eq:momentum}) describes the case where the impurity-boson interaction is switched on adiabatically slowly such that no excitations are generated.

	We define the energy $E_p$ of the impurity dressed by the cloud of bosons as a difference of the ground state energy of the Hamiltonian (\ref{eq: HLee-Low}) at a given total momentum $p$ and its ground state energy at zero momentum in the absence of the impurity. We evaluate the polaron energy $E_p$ to be
	\begin{align}
		\frac{E_{p}}{g n_0}={}&\frac{2}{3}b^3\frac{\sqrt{1+m/M} }{\sqrt{\gamma }}\left[1-\tanh
		^3(x_0)\right]+\frac{MV_f^2}{2g n_0}\notag\\ &+\frac{1}{3} b^3
		\frac{\sqrt{1+m/M} }{\sqrt{\gamma }} \left[\tanh ^3(x_0)-3 \tanh
		(x_0)+2\right],
		\label{eq:PolaronEnergy}
	\end{align}
	as in the case of a repulsively interacting impurity \cite{AnnalsKamenev,Quench}. Here $V_f$ satisfies Eq.~(\ref{eq:momentum}) and $x_0$ obeys \eq{eq:xo}. Using the path-integral approach, the energy dispersion has been evaluated in Ref.~\cite{Pastukhov2019} in a somewhat cumbersome integral form. Evaluating the latter, we have verified that it gives the polaron energy in agreement with \eq{eq:PolaronEnergy}.
	From \eq{eq:momentum} follows that changing the momentum as $p\to p\pm 2\pi\hbar n_0$, $V_f$ and $x_0$ remain unaltered. As a result, the energy (\ref{eq:PolaronEnergy}) is a periodic function of momentum, with the period $2\pi\hbar n_0$.

	\begin{figure}
		\includegraphics[width=0.95\columnwidth]{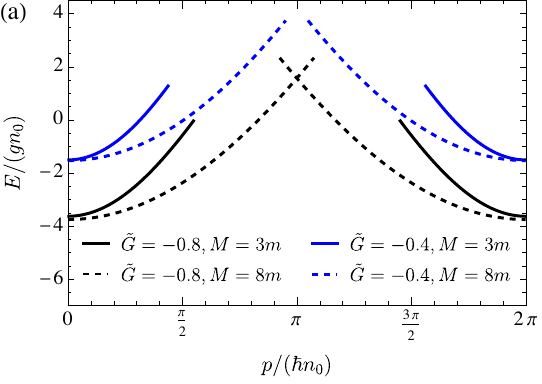}  
		\includegraphics[width=0.95\columnwidth]{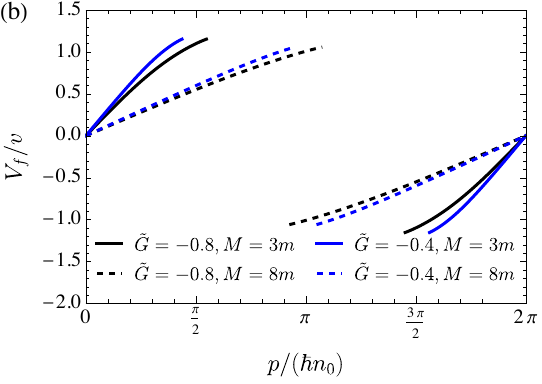} 
		 \caption{(a) Ground-state energy dispersion (\ref{eq:PolaronEnergy}), and (b) final impurity velocity (\ref{eq:momentum}) of the ground state (\ref{eq:MFsolution}) as a function of the system momentum for different impurity masses and coupling constants. Here $\gamma=0.1$. Note that $M=8m>M_c$ for $\tilde{G}=-0.8$.}
		\label{fig2}
	\end{figure}
	
Figure \ref{fig2} shows the energy dispersion (\ref{eq:PolaronEnergy}) and the final impurity velocity (\ref{eq:momentum}) as a function of the total system momentum for different coupling constants $\tilde{G}$ and impurity masses. Contrary to the repulsive case \cite{Quench}, the physical solution (\ref{eq:MFsolution}) is allowed only in a certain interval of momenta in a general case. The latter is given by \eq{eq:momentum}, where the final velocity takes values in the interval $(-v_c,v_c)$. The reason for the existence of forbidden momenta is the absence of a second physical solution of \eq{eq:xo} for a given $|V_f|<v_c$. Moreover, the energies of the two other solutions of \eq{eq:xo} do not merge with the physical one, see App.~\ref{sec:Appendix}. However, we will show in Sec.~\ref{sec:relaxation} that the stationary boson density profile around the impurity obtained after a sudden quench of the impurity-boson interaction, for an \emph{arbitrary} system momentum, is given by the physical solution \eq{eq:MFsolution}, with a final velocity $V_f$ that does not satisfy \eq{eq:momentum}. The reason for the latter is the emission of excitations in the form of dispersive density shock waves, density waves and solitons that take away some momentum. This implies that such emission is requisite for the system to access state (\ref{eq:MFsolution}) when the initial momentum lies within the forbidden region.
	
Increasing the impurity mass or its coupling constant, the range of momenta where the analytic solution (\ref{eq:MFsolution}) exists increases, and the two energy branches approach each other, see Fig.~\ref{fig2}(b). Thus, as for a repulsive impurity potential \cite{lamacraft2009dispersion,PhysRevLett.108.207001}, one can define the critical mass $M_c$ when the two energy branches touch each other. As a result, the cusps appear in the ground-state energy at momenta $p=(2m+1)\pi \hbar n_0$, with $m$ being an integer. Then, \eq{eq:MFsolution} is the ground state at all momenta. For $M>M_c$, the energy branches cross each other leading to a rich energy landscape and the appearance of metastable states. 

Note that for $M>M_c$ the critical velocity is not given by $v_c$ once the quantum fluctuations are taken into account. The velocity $v_c$ gives the final impurity velocity at the end of the energy branch. The latter describes a metastable state. Quantum fluctuations allow for tunneling from the metastable state into the ground state, and the system dissipates energy while relaxing into the ground state. Thus the critical velocity, defined as the velocity below which an impurity moves without any friction at zero temperature, is $\tilde{v}_c$ and is given by the final velocity evaluated at $p=\pi \hbar n_0$ using \eq{eq:momentum}, as in the case of a repulsive impurity \cite{PhysRevLett.108.207001}. The critical velocity $\tilde{v}_c$ decreases as $M$ increases and becomes zero in the limit of an infinitely heavy impurity.

Comparison of the binding energy, $E_{p=0}$, with the diffusion Monte Carlo results of Ref.~\cite{Grusdt_2017} shows a very good agreement  in a wide interval of impurity-boson interactions, while the agreement for the effective impurity mass is very good at weak to moderate interactions only \cite{PhysRevResearch.2.033142}\footnote{The result for the zero momentum polaron energy given in Ref.~\cite{PhysRevResearch.2.033142} agrees with ours after correction of the typo in the definition of their parameter $y$. It should read as $y=z-\sqrt{1+z^2}$, with $z=2\sqrt{2} n_0 \tilde{\xi}/\eta$.}.	
At $\gamma=0.02$, the difference between the mean-field results and the Monte Carlo results of Ref.~\cite{Parisi2017} for the effective mass \cite{Pastukhov2019} and the contact parameter  \cite{10.21468/SciPostPhys.11.1.008}  is noticeable at weak impurity-boson interactions, while at $\gamma=0.2$ the agreement is very good in a wide interval of impurity-boson interactions  \cite{Pastukhov2019}. 
Moreover, these comparisons consider $M\leq m$, while quantum fluctuations become less relevant for heavier impurities.
	
Note that in the case of a strong attractive coupling, $|\tilde{G}|\gg 1$, the number of bosons gathered around the impurity is negligible with respect to the number of bosons in the remaining part of the system as long as 
$|\tilde{G}|\ll L{\sqrt{m g n_0}}/{\hbar}$. This inequality gives also the condition under which the expansion (\ref{eq:muEv}) is valid. However,  the density at the impurity position grows with $\tilde{G}$ as $n(0)= n_0 \tilde{G}^2/(1+m/M)$, and the polaron energy (\ref{eq:PolaronEnergy}) behaves as $E_p\propto\tilde{G}^3 g n_0$. It is expected that a modeling of the impurity-boson interaction with a finite-range potential instead of the contact one will cure this problem \cite{gqd6-t881}. Namely, it was shown that a finite-range impurity potential leads to a finite  zero-momentum polaron energy in the strong coupling limit \cite{gqd6-t881}. Also, at strong coupling, a limited number of potentially bound bosons \cite{gunn1988,Kolomeisky,EnssPhysRevResearch.2.032011} may become relevant. In the following sections, the phenomena we are interested in do not require a strong coupling, and we consider a weak to moderate coupling constant $\tilde{G}$.

	\section{Post-quench dynamics of polaron formation and relaxation \label{sec:relaxation}}

	\begin{figure*}
		\includegraphics[width=0.9\columnwidth]{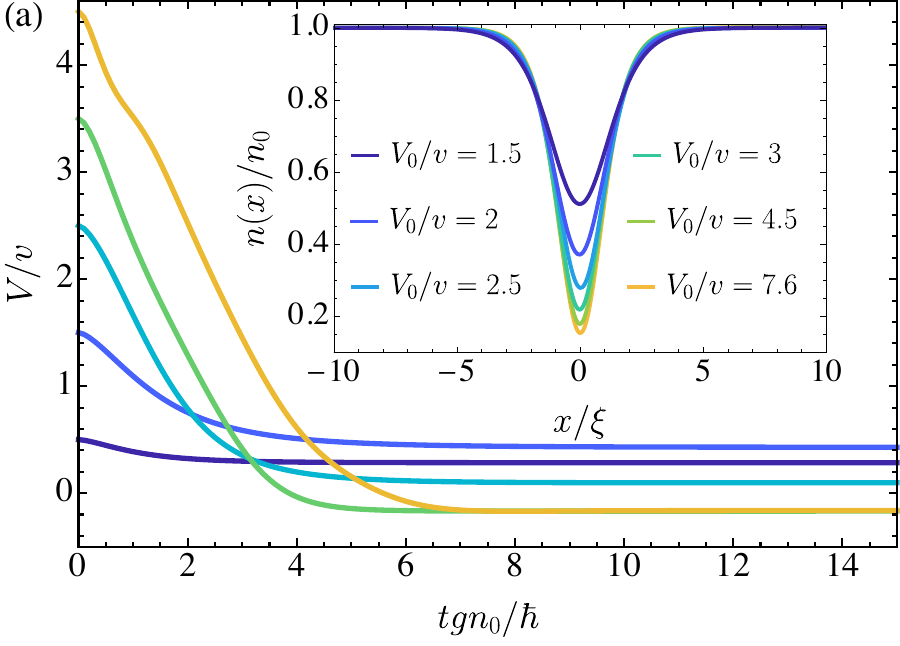}  \phantom{aaaa}
		\includegraphics[width=0.95\columnwidth]{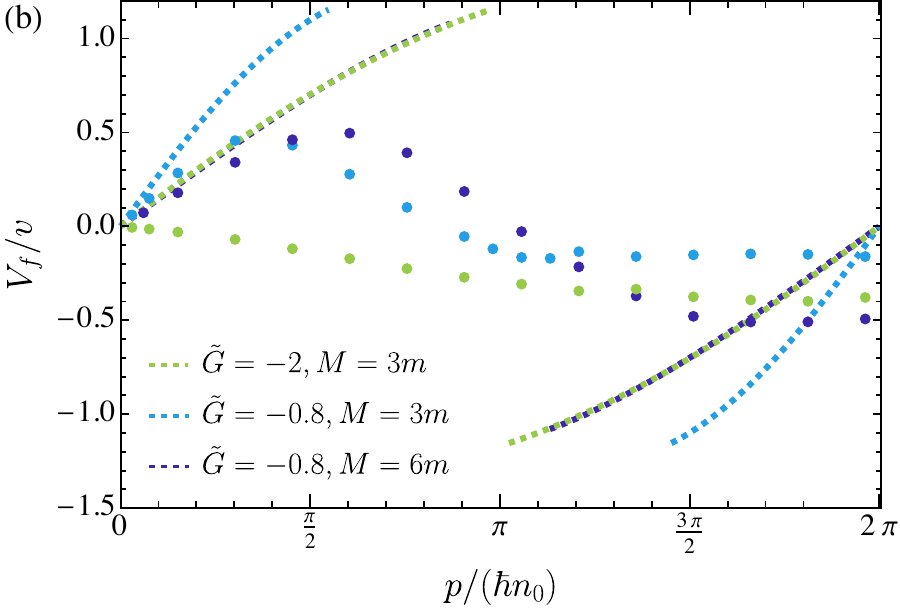}  
		\caption{(a) Time-evolution of the impurity velocity for $M=3m$ and $\tilde{G}=-0.8$ for different initial velocities. Here $\gamma = 0.1$. (\textit{inset}) The density profile of emitted soliton for different initial impurity velocities $V_0$ for the aforementioned parameters. (b) Final impurity velocity $V_f$ as a function of the initial impurity momentum $p=M V_0$ for three different sets of parameters. Here, the dotted lines denote the analytic expression (\ref{eq:momentum}), while the points denote the numerically obtained values of $V_f$. One can express $p/\hbar n_0=(V_0/v) (M\sqrt{\gamma}/m)$.}
		\label{fig4}
	\end{figure*}

	In this section, we study the system time evolution after a sudden injection of a finite-momentum impurity into the background bosons. At initial time the impurity is free, while the bosons are in their zero-momentum ground state $\Psi_0(x,0)=\sqrt{n_0}$. 
We solve numerically the time-dependent Gross-Pitaevskii like equation (\ref{eq:mean-field1}) with periodic boundary conditions by implementing a conservative finite-difference scheme in a fully implicit manner \cite{GPE_discretization}. 
A first-order upwind scheme \cite{upwind_scheme} is used to discretize the drift term, maintaining numerical stability and ensuring physical causality by accounting for the direction of the impurity velocity.
The obtained system of non-linear equations is solved in an iterative fashion \cite{GPE_discretization}. The system size is taken to be sufficiently long, such that the emitted density waves do not reach the boundaries of the system during the reported relaxation. 
	
	After the immersion, the impurity slows down on the time-scale of several $\hbar/g n_0$, as shown in  Fig.~\ref{fig4}(a). The higher the initial velocity is, the longer the impurity relaxation time is. The impurity transfers a part of its momentum to the bosonic bath by triggering the emission of dispersive density shock waves \cite{PhysRevA.69.063605,PhysRevA.74.023623}, see Fig.~\ref{fig1}(a). Due to the attractive nature of the impurity-bath interaction, the bosons near the impurity gather around it forming a peak in the density. For a sufficiently high initial impurity velocity, a depletion hole separates from the density shock wave, taking the form of a gray soliton, as illustrated in Fig.~\ref{fig1}(b). As time progresses, the shock-wave fronts and the soliton move away from the impurity location, the impurity velocity reaches a final stationary value smaller than the critical velocity and locally the ground state (\ref{eq:MFsolution}) with $n(0)>n_0$ is established. As a hallmark of the zero temperature bath superfluidity, the impurity final velocity is nonzero in a general case.
	
	We emphasize that \eq{eq:MFsolution} with $n(0)>n_0$, for a numerically obtained $V_f$, describes the boson density peak around the impurity in obtained stationary states for \emph{all} values of the initial impurity momentum $p=M V_0$, including the forbidden momentum sector. The latter is given by values of $p$ different from those obtained by \eq{eq:momentum} with $V_f$ satisfying $|V_f|\leq v_c$. However, the final impurity velocity is not given by \eq{eq:momentum}, see Fig.~\ref{fig4}(b), since some momentum is transmitted to the density waves and the soliton, and these excitations are not taken into account by the analytic solution (\ref{eq:MFsolution}). Equation (\ref{eq:momentum}) is valid in the case where the impurity-boson interaction is turned on adiabatically slowly such that no excitations are generated.

	\begin{figure}
		\includegraphics[width=0.9\linewidth]{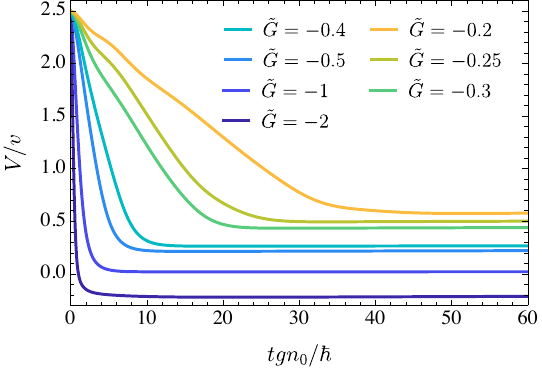}  
		\caption{Time evolution of the impurity velocity for $M=3m$ and $V_0=2.5v$ for different dimensionless impurity-bath coupling $\tilde{G}$. Here $\gamma = 0.1$.}
		\label{fig5}
	\end{figure}
	
	Increasing the initial impurity velocity $V_0$ above some threshold, the final velocity decays and more energetic solitons are emitted, as illustrated in Fig.~\ref{fig4}. As a result, the maximal soliton depletion increases with  $V_0$. However, both the soliton depth and the final impurity velocity show a tendency of smooth saturation into a constant value for a sufficiently high $V_0$.  As a result, the system remains locally in the same stationary state as $V_0$ is further increased, while the energy and momentum carried by the dispersive shock waves increase. A similar dynamical crossover was reported for a repulsive impurity-boson interaction \cite{Quench}. 
	
	For stronger impurity-bath coupling, the deviation between numerically obtained values of  $V_f$ and those following from \eq{eq:momentum} becomes more pronounced even at low $p$, due to the emission of bigger-amplitude shock waves, see Fig.~\ref{fig4}(b). For the case of $\tilde{G}=-2$ and $M=3m$ shown in Fig.~\ref{fig4}b, for $V_0>2.5v$, the impurity dynamics changes drastically and a new dynamical regime occurs, which will be discussed in Sec.~\ref{sec:resonant}. There, the impurity exhibits transient temporal velocity oscillations before reaching the stationary state (\ref{eq:MFsolution}).

	The impurity time evolution as a function of the dimensionless impurity coupling constant $\tilde{G}$ is shown in Fig.~\ref{fig5} for $M=3m$ and $V_0=2.5v$. Here, the impurity mass is below the critical one, $M< M_c(\tilde{G})$, for all the values of $\tilde{G}$. The impurity relaxation time decreases as $\tilde{G}$ increases. In the parameter region shown in Fig.~\ref{fig5}, the final impurity velocity also decreases with $\tilde{G}$, and finally, for a very strong coupling, the impurity changes the direction of motion with respect to the initial one. 
	
\begin{figure}
\includegraphics[width=0.9\linewidth]{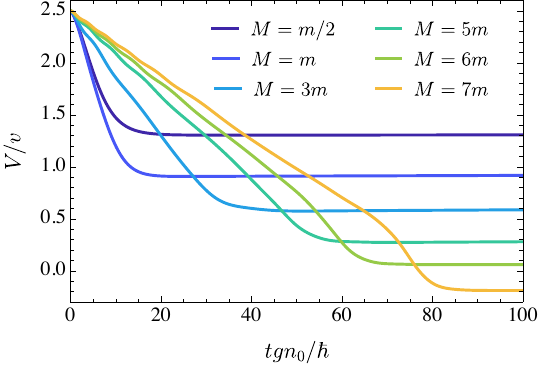}  
\caption{Time evolution of the impurity velocity for $\tilde{G}=-0.2$ and $V_0=2.5v$ for different impurity masses. Here $\gamma = 0.1$. }
\label{fig6}
\end{figure}

	We study the relaxation dynamics for different values of the impurity mass $M<M_c$ in Fig.~\ref{fig6}, while keeping other parameters fixed, $\tilde{G}=-0.2$ and $V_0=2.5v$. For heavier impurities, the deceleration is slower, leading to a longer relaxation time. For considered parameters, Fig.~\ref{fig6} shows the final impurity velocity that decreases with $M$. As a result, the energy transferred to the bath increases with $M$, and  the emitted solitons become more energetic. Sufficiently heavy impurity changes the initial direction of motion, reaching a negative final velocity. Carefully looking at Fig.~\ref{fig4}(b), one sees that the final velocity can also become higher by increasing the ratio $M/m$ for the same initial velocity and $\tilde{G}$, see $V_0=\pi v/(10\sqrt{\gamma})$ ($p=3\pi \hbar n_0/10$ for $M=3m$). The reason is the nonmonotonic dependence of $V_f$ on $V_0$. 
	
\section{Post-quench dynamics using an alternative equation of motion\label{sec:alternative}}

Note that \eq{eq:mean-field1} gives the modified sound velocity $v\sqrt{1+m/M}$ even at vanishingly small impurity-boson interaction strength. Thus, although all the obtained final impurity velocities are smaller than the critical velocity $v_c$, a light impurity can have the final velocity higher than the sound velocity $v$, since $V_f\leq v_c=v\sqrt{1+m/M}$, see Fig.~\ref{fig6}. This problem does not arise if the $1/M$ contribution in the first term in \eq{eq:mean-field1} is absent, leading to
\begin{align}
		i\hbar \partial_t{{\Psi_0}(x,t)}=\Bigg[&-\frac{\hbar^2}{2m}\partial_x^2+g |{\Psi_0}(x,t)|^2\notag\\&+G\delta(x)+i\hbar V(t)\partial_x\Bigg] {\Psi_0}(x,t),
		\label{eq:mean-field2}
\end{align}
with $V(t)$ given by \eq{eq:Vimp}. Equation (\ref{eq:mean-field2}) gives the correct sound velocity in the laboratory frame $v=\sqrt{g n_0/m}$ \cite{pitaevskii_bose-einstein_2003}.

The alternative equation of motion (\ref{eq:mean-field2}) is studied in Ref.~\cite{Quench}. There, the comparison of the polaron energy obtained within the mean-field approach with the exact result for the Yang-Gaudin model is performed, showing that \eq{eq:mean-field2} captures the underlying physics more accurately than \eq{eq:mean-field1} for nonzero system momenta.

\begin{figure}
		\includegraphics[width=0.9\columnwidth]{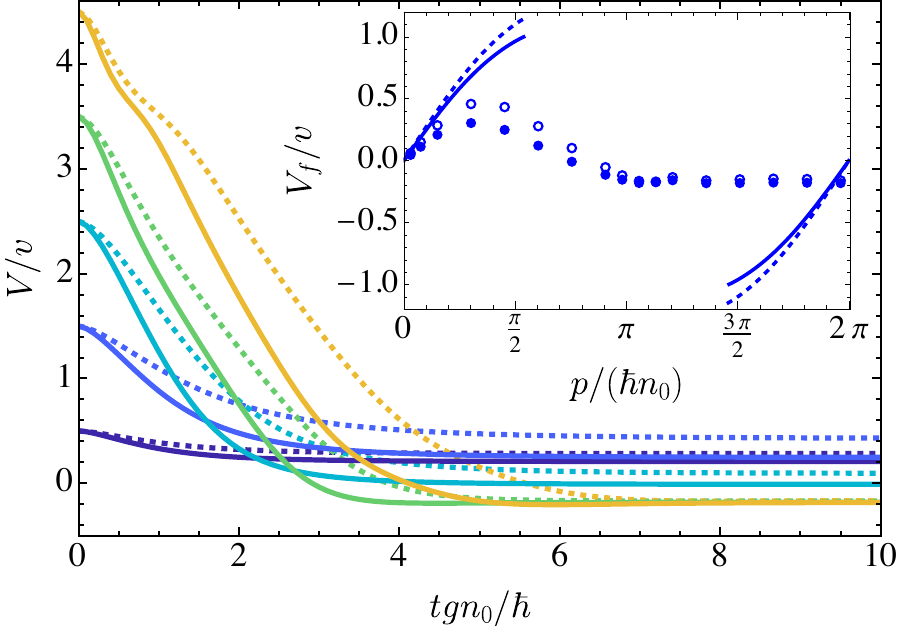} 
		\caption{Impurity velocity as a function of time for $\tilde{G}=-0.8$, $M=3m$, $\gamma=0.1$ for different initial velocities, obtained using \eq{eq:mean-field1} (the dashed line) and the alternative equation of motion (the solid line). The inset shows the final impurity velocity as a function of initial impurity momentum for the same parameters. The lines and the circles represent the analytical prediction in the case of the adiabatic turning-on of the impurity-boson interaction (the {dashed} and the {solid} line show the \eq{eq:momentum} and its modified version for the alternative equation of motion, respectively) and the numerically obtained values (the {hollow} circles for the standard and the {filled} ones for the alternative  equation), respectively.}
		\label{fig:new_GPE}
	\end{figure}

To characterize the stationary state of \eq{eq:mean-field2}, one should replace $1+m/M$ by $1$ in all the equations in Sec.~\ref{sec:Apolaron}. The differences between the two approaches are thus expected to be visible for not too heavy impurities. 
Using \eq{eq:mean-field2}, we examine the post-quench dynamics in Fig.~\ref{fig:new_GPE} and compare it to the previously obtained results shown in Fig.~\ref{fig4}(a). Now, the impurity relaxation towards a stationary state becomes faster, and the obtained final velocities differ. In the regime where $V_f$ saturates, the differences in $V_f$ become smaller.

	\section{Oscillations}\label{sec:resonant}

	\subsection{Physical mechanism\label{sec:Mechanism}}
	
	\begin{figure*}
		\includegraphics[width=\columnwidth]{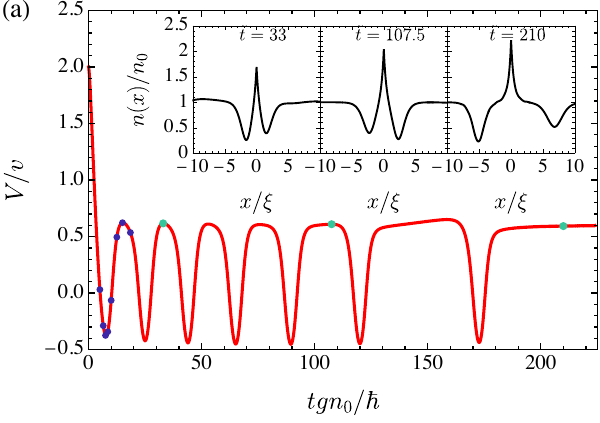}
		\phantom{aaa}
		\includegraphics[width=\columnwidth]{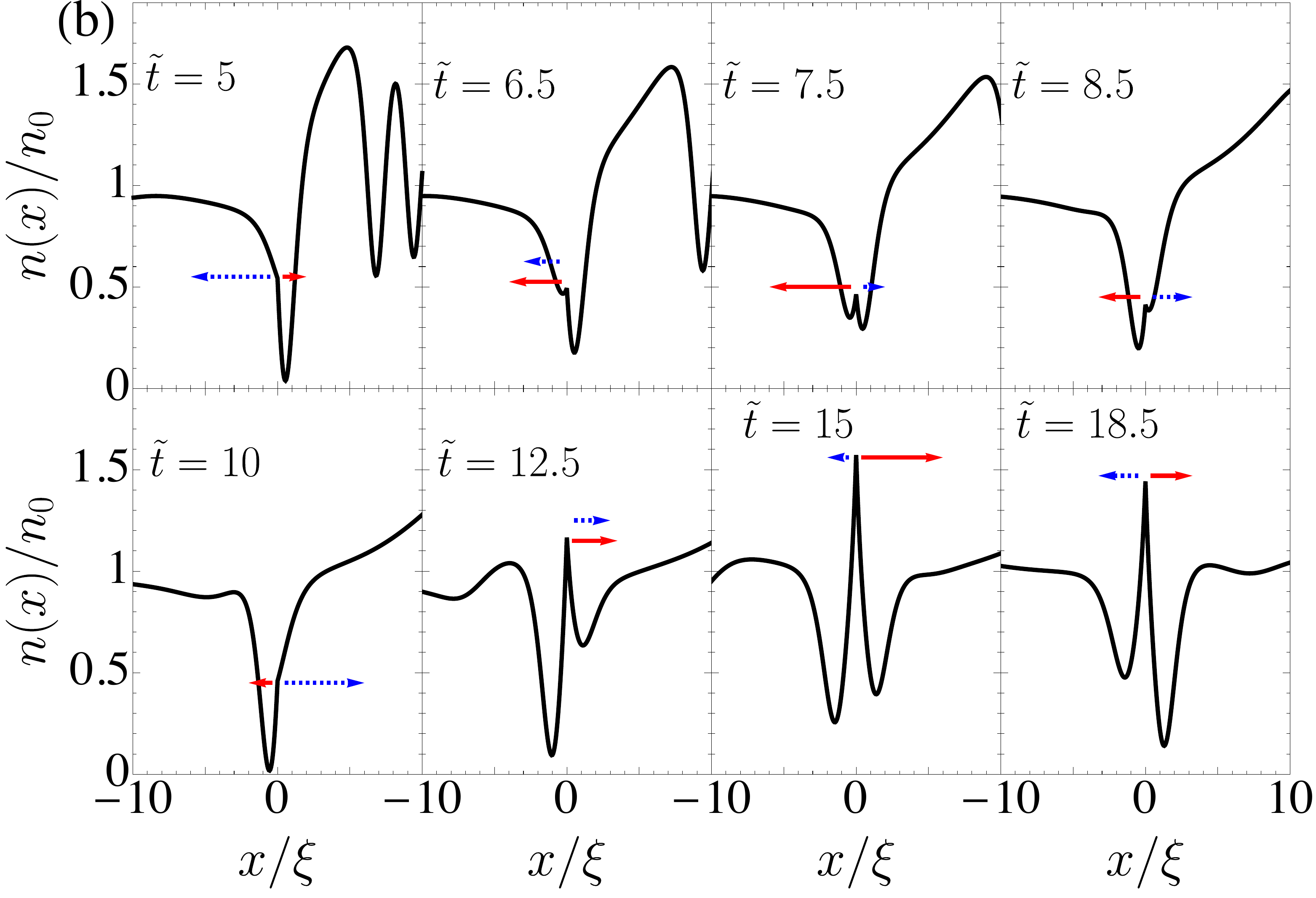}
		\caption{(a) Impurity velocity as a function of time for $\tilde{G}=-0.8$, $M=9m$, $\gamma=0.1$, and  $V_0=2v$. The inset shows the boson density in the frame co-moving with
the impurity at different times, marked with light green points in the velocity time dependence. In the inset, the $x$ axis represents the dimensionless length $x/\xi$, and the dimensionless time is defined as $\tilde{t}=t gn_0/\hbar$. (b) Boson density $n(x)/n_0$ in the frame co-moving with
the impurity for the aforementioned parameters shown over one period of oscillations at the time-points marked by dark blue points on the velocity curve in (a). The red solid and the blue dashed arrows denote the
direction of the dimensionless impurity velocity $V/v$ and the dimensionless force acting on the impurity $F/(gn_0^2)$ at the specific timeframes in the laboratory frame, respectively. The lengths of the arrows are proportional to their magnitudes.}
		\label{Resonant}
	\end{figure*}
	
	In this section, we study the out-of-equilibrium dynamics of a fast impurity with a mass greater than or close to the critical one, where a novel dynamical regime takes place. The latter is illustrated in Fig.~\ref{Resonant}(a) for $\tilde{G}=-0.8$, $M=9m$, and  $V_0=2v$. After a rapid drop, the impurity velocity exhibits undamped long-lived oscillations in time, before reaching a stationary state. 
	
	In order to explain the underlying mechanism, we study the accompanying time evolution of the boson density in the vicinity of the impurity in Fig.~\ref{Resonant}(b). Note that the force exerted by the bosons on the impurity is given by the gradient of the potential, and reads as
	\begin{align}\label{eq:friction}
		F(t)=&\int \mathrm{d} x \Psi_0^*(x,t) \partial_x\left[G \delta(x)\right] \Psi_0(x,t)\notag\\=&-\frac{G}{2}\left[\partial_x n(x,t)|_{x=0^-}+\partial_x n(x,t)|_{x=0^+} \right].
	\end{align}
	In Fig.~\ref{Resonant}(b), the red solid and the blue dashed arrows denote the vectors of the dimensionless impurity velocity $V/v$ and the dimensionless force, $F/(gn_0^2)$,  acting on the impurity, respectively. After the injection into the condensate, the impurity decelerates rapidly while emitting dispersive density  shock waves, as in a general case studied in Sec.~\ref{sec:relaxation}. Meanwhile, the boson density peak builds up at the impurity position. The impurity slows down and reaches zero velocity. The density profile close to this time, shown in Fig.~\ref{Resonant}(b) at $\tilde{t}=t g n_0/\hbar=5$,  reveals that the heavy and fast impurity has created an almost complete depletion in front of it.  The interference between the density peak and this depletion area makes the peak nearly invisible. As time passes, the shock wave moves away from the impurity. In contrast to the scenario of Sec.~\ref{sec:relaxation}, the depletion cloud here does not separate from the impurity, and instead forms an oscillating state with the density peak and the impurity.
	The steep boson density slope caused by the depletion that remains attached to the impurity, in front of it, results in a negative force (\ref{eq:friction}) acting on the impurity. Thus, the impurity changes the direction of motion and accelerates, building up the density peak at its position, as well as a new depletion hole positioned in front of it, as shown at $\tilde{t}=6.5$. Simultaneously, the previous depletion cloud, now situated behind the impurity, diminishes. As a result, the magnitude of the force $|F(t)|$ decays. The impurity reaches its maximal negative velocity and the slope of the density profile becomes symmetric around the impurity, leading to a vanishing friction force, see the density profile at $\tilde{t}=7.5$ that is close to this characteristic time. Afterwards, the depletion hole in front of the impurity continues to grow, while the one behind it diminishes further, as shown at $\tilde{t}=8.5$. As a result of this asymmetry in the boson density, the friction force changes its direction and the impurity slows down, resulting in a less pronounced density peak. The increasing boson density asymmetry leads to a continuous increase of the friction force in time. The latter reaches its maximal value, while the impurity velocity arrives at zero value, see Fig.~\ref{Resonant}(b) at $\tilde{t}=10$. The impurity again changes its direction of motion, and this process repeats. However, the magnitude of the maximal positive impurity velocity exceeds the maximal negative one, and the density shows a more prominent peak at the impurity position at $\tilde{t}=12.5$, $15$ and $18.5$. The above described process can be understood as a collision of the depletion cloud and the peak, where the latter gets trapped and oscillates inside the depletion hole.

\begin{figure}
		\includegraphics[width=\columnwidth]{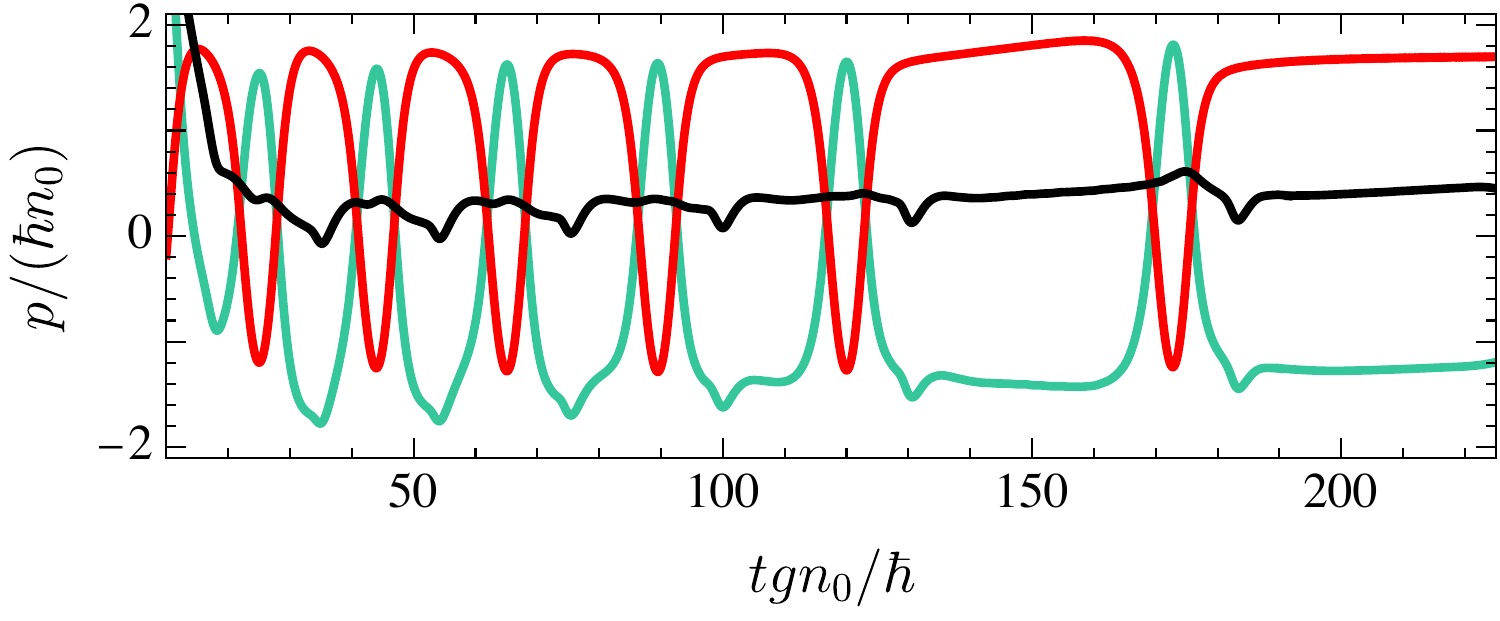}
		\caption{Momentum of bosons  $p_b$ situated in the vicinity of the impurity in the interval $x\in(-15\xi,15\xi)$ and the impurity momentum $p_i$ are shown by the light green and the red line, respectively. The black line represents $p_b+p_i$. Here $\tilde{G}=-0.8$, $M=9m$, $V_0=2v$, and $\gamma=0.1$.}
		\label{Energy-momentum}
	\end{figure}

	The evolution of the impurity momentum $p_i(t)=MV(t)$ and the momentum of bosons situated around the impurity $p_b(t)=-i\hbar \int_{-W}^{W} dx \Psi_0^*(x,t) \partial_x \Psi_0(x,t) $ is shown in Fig.~\ref{Energy-momentum}. Here, the half-width of the interval is $W=15 \xi$, such that it encompasses the impurity and the depletion hole during the oscillations. The momenta of the impurity and the neighbouring bosons have opposite signs and oscillate in anti-phase. The total momentum, $p_b+p_i$, also exhibits oscillations, but with a very small amplitude. Moreover, the maximum of the boson density peak is always situated at the impurity position and does not oscillate around it. Note that the oscillations trigger emission of density waves propagating in the remaining part of the system, resulting in a decrease of the energy of the subsystem composed of the impurity and the surrounding bosons.

	Each time the system passes through the state with the maximal impurity velocity, the two depletion clouds positioned on the opposite sides of the impurity become more separated from the impurity, and thus less influencing the density profile at the impurity position and the resulting force. As a result, the density peak realized in this state grows in time, as shown in the inset of Fig.~\ref{Resonant}(a). Moreover, the friction force remains approximately zero for a longer time during which the system stays in this maximal-$V$ state. Finally, the repulsion between the depletion clouds and the impurity and the depletion-depletion repulsion win and the two depletion holes separate and move away from the impurity in the opposite directions, taking the form of grey solitons. As a result, the stationary configuration (\ref{eq:MFsolution}) is realised locally, and the density peak moves together with the impurity. The final impurity velocity is close to the maximal positive velocity of intermediate time oscillations.	
	
We stress that the absolute value of the impurity velocity during the oscillations is smaller than the critical velocity $\tilde{v}_c$ introduced in Sec.~\ref{sec:Apolaron}. Thus, the impurity motion remains disipationless even in the presence of quantum fluctuations. This is the case for all the shown examples in the following subsections.

\subsection{Dependence on the initial impurity velocity \label{sec:ResonantV}}
	
	\begin{figure*}
		\includegraphics[width=\linewidth]{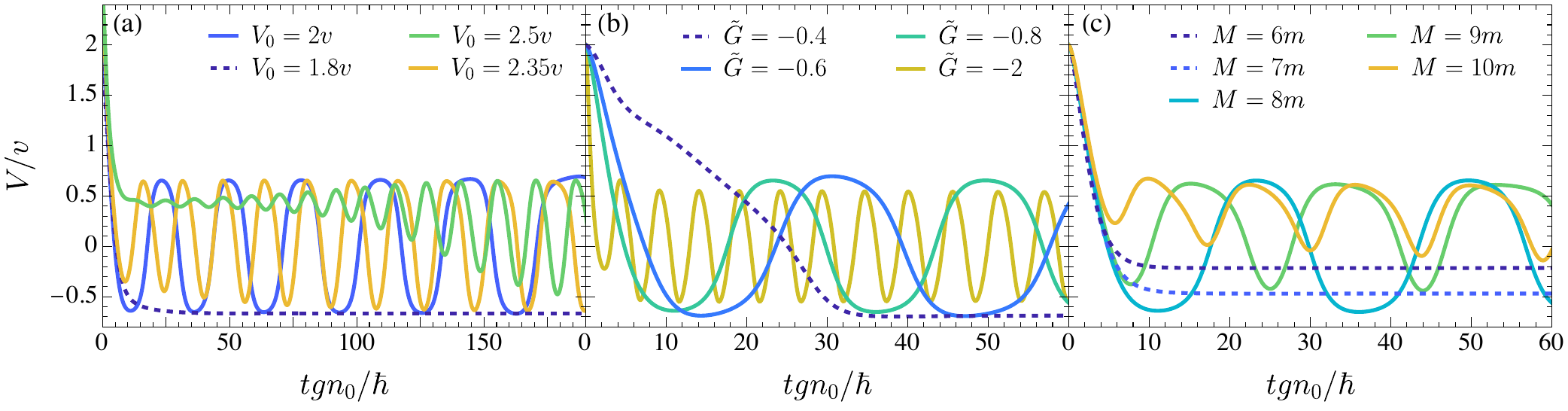}
		\caption{Impurity velocity as a function of time (a) for $\tilde{G}=-0.8$ and $M=8m$ for different values of the initial impurity velocity, (b) for $M=8m$ and $V_0=2v$ for different impurity-bath coupling $\tilde{G}$, and (c) for $\tilde{G}=-0.8$ and $V_0=2v$ for different impurity masses. At all figures $\gamma=0.1$.}
		\label{ResonantVGM}
	\end{figure*}
	
	The effect of the initial impurity velocity $V_0$ on the dynamics and the formation of the oscillating state is illustrated in Fig.~\ref{ResonantVGM}(a). The time evolution of the impurity velocity for different values of $V_0$ is shown for $\tilde{G}=-0.8$, $M=8m$ and $\gamma=0.1$. For $V_0\leq 1.8 v$, the impurity relaxes to a stationary state in a standard manner, as described in Sec.~\ref{sec:relaxation}. For $V_0=2v$, the oscillating state characterized in Sec.~\ref{sec:Mechanism} is realized. In this instance, however, the velocity oscillations are symmetric around $V=0$, such that $V_{\mathrm{max}}\approx -V_{\mathrm{min}}$. 
At higher initial impurity velocity, we see another aspect of this phenomenon where the system enters into oscillations while the density peak at the impurity position is not fully formed. While this trend was also present for the parameters of Fig.~\ref{Resonant}, the effect is significantly more pronounced for the current parameters.
As a result, the impurity velocity exhibits a progressive increase of the oscillation amplitude in time, while the density peak forms as it oscillates. These oscillations then stabilize, i.e., the oscillation amplitude becomes constant, before reaching a stationary case. This dynamical regime is shown  in Fig.~\ref{ResonantVGM}(a) for $V_0=2.35v$ and $V_0=2.5 v$. Moreover, Fig.~\ref{ResonantVGM}(a) suggests that the oscillation frequency increases as $V_0$ is increased.

	\subsection{Dependence on the strength of the impurity coupling  \label{sec:ResonantG}}

	Next, we examine the influence of $\tilde{G}$ on the oscillation. Figure \ref{ResonantVGM}(b) shows the impurity velocity time evolution for different values of $\tilde{G}$ for $M=8m$ and $V_0=2v$. 
	At $\tilde{G}=-0.4$, the impurity does not exhibit the velocity oscillations for the aforementioned parameters. In this case, the impurity mass is below, but in the close vicinity of the critical mass. The shock wave and the complete depletion cloud are formed at $\tilde{t}=5$, while only a partially formed density peak gets captured inside the depletion cloud till $\tilde{t}=30$.  
	Then the depletion gets split into two parts that move away from the impurity. Increasing $\tilde{G}$, the process of the peak formation is more rapid, and the impurity enters into the oscillating state. 
	By making the impurity-bath coupling more attractive, the absolute value of the force acting on the impurity increases, $F \sim G $,  leading to a higher-frequency and smaller-amplitude velocity oscillations. For the parameters shown in Fig.~\ref{ResonantVGM}(b), the oscillations remain symmetric around $V=0$, with $V_{\mathrm{min}}\approx - V_{\mathrm{max}}$. The amplitude of density waves triggered by the impurity oscillations increases with the absolute value of the impurity-bath coupling, leading to a bigger rate of  energy decrease of the subsystem consisting of the combined structure of the density peak with the depletion holes and the impurity. We stress that the lifetime of the oscillating state increases by increasing $|\tilde{G}|$. For example, a change of $\tilde{G}$ from $-0.6$ to $-0.8$ leads to an approximately $70\%$ longer lifetime.

	\subsection{Dependence on the impurity mass \label{sec:ResonantM}}

\begin{figure}
		\includegraphics[width=0.95\columnwidth]{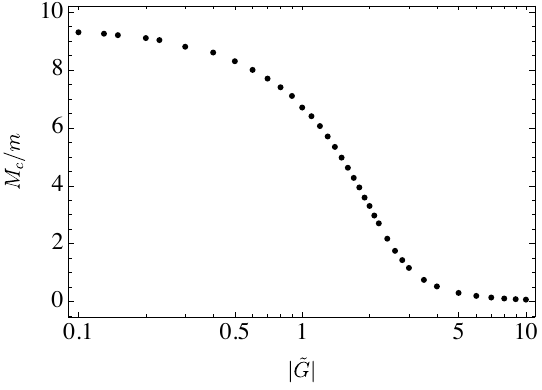}
		\caption{Critical mass as a function of the impurity-boson interaction strength. Here, $\gamma=0.1$.}
		\label{fig:Mc}
\end{figure}

In order to highlight the parameter region where the oscillations occur, we show the critical mass as a function of the impurity-boson interaction strength in Fig.~\ref{fig:Mc}. We find that the critical mass reads as
\begin{align}
M_c(\tilde{G}=0^-)=m \frac{2\pi^2}{\gamma+\sqrt{4\pi^2\gamma+\gamma^2}},
\end{align}
at $\tilde{G}=0^-$. At stronger attraction, the critical mass is smaller. It  monotonically decreases with $|\tilde{G}|$. 
	
The effect of the impurity mass $M$ on the oscillations is shown in Fig.~\ref{ResonantVGM}(c) for $\tilde{G}=-0.8$ and $V_0=2v$. For lighter impurities, $M=6m$ and $7m$, the oscillations are absent and the system follows the scenario of Sec.~\ref{sec:relaxation}. Note that these values of the impurity mass are smaller than $M_c$. Increasing $M$, the oscillations occur. As the impurity becomes heavier $M>8m$, $V_{\text{max}}$ almost remains the same, whereas $V_{\text{min}}$ increases rapidly, decreasing the oscillation amplitude. 
This asymmetry manifests also in the boson density at minima and maxima of $V(t)$, with the boson density peak being smaller at $V_{\text{min}}$ than at $V_{\text{max}}$. Moreover, the maximal value of the force (\ref{eq:friction}) felt by the impurity increases slightly with $M$. However, the amplitude of the impurity velocity oscillations decreases rapidly with $M$, thus leading to an increase of the oscillation frequency. 
	
	For $M=10m$, the oscillation amplitude increases in time in the initial stage of the oscillations because the peak in the boson density at the impurity position is still forming. A heavier impurity produces also more important density waves during oscillations, resulting in a higher rate of energy decay of the subsystem composed of the boson density peak, the impurity and the depletion cloud. Finally, note that the lifetime of the oscillations at $M=8m$ is approximately $95\%$ longer than for $M=9m$. The latter is shown in Fig.~\ref{Resonant}(a). 
	
	\begin{figure}[t]
		\includegraphics[width=0.9\linewidth]{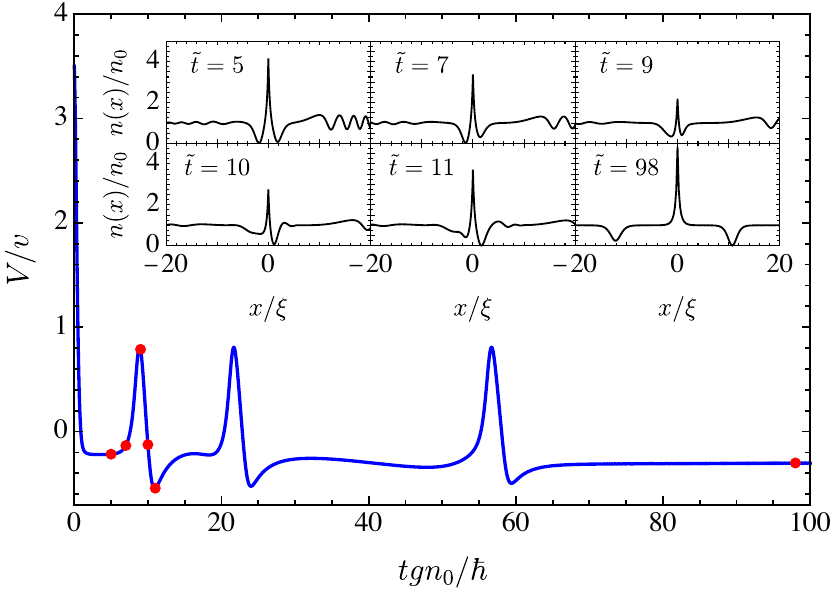} 
		\caption{Time evolution of the impurity velocity for $M=3m$, $V_0=3.5v$ and $\tilde{G}=-2$. Here $\gamma = 0.1$. The inset shows the time evolution of the boson density around the impurity.}
		\label{fig:strongG}
	\end{figure}
	
	We stress that the oscillating state can also be formed for an impurity lighter than the critical one.
For example, for the parameters of Fig.~\ref{ResonantVGM}(c), the impurity with a mass $M=7m<M_c$ does not oscillate at $V_0=2$, but increasing its initial velocity to $V_0=2.5$ the above described oscillating state is realised. Another example is a moderate $\tilde{G}$, e.g., with $\tilde{G}=-2$, $M=3m<M_c$, $V_0\geq 3v$ and $\gamma=0.1$. The dynamics for $V_0=3.5v$ is shown in Fig.~\ref{fig:strongG}. Here, the shape of the oscillations is quite different. However, the underlying mechanism is the same, the transient localization of the depletion hole around the impurity position, as illustrated in the inset of Fig.~\ref{fig:strongG}. In Fig.~\ref{fig4}(b), we present the final velocity as a function of the initial one. We observe that the lifetime of the oscillations is a non-monotonic function of $V_0$ for these parameters. 
	
\subsection{Repulsively interacting impurity}
	
\begin{figure}
		\includegraphics[width=0.9\columnwidth]{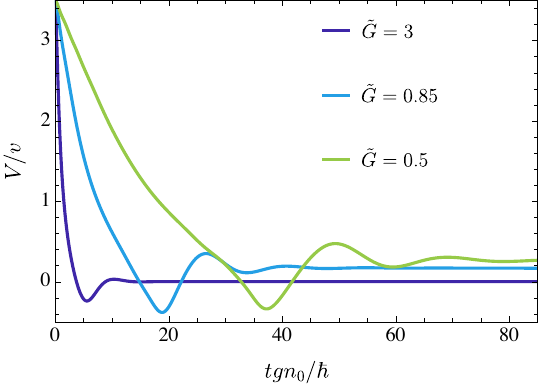} 
		\caption{Time evolution of the impurity velocity for $M=8m$ and $V_0=3.5v$ for different strengths of repulsive impurity-boson interaction. Here $\gamma = 0.1$.}
		\label{fig:repulsive}
\end{figure}

Here, we make a comparison with the relaxation of a repulsively interacting impurity in a system of a 1d bosons. In the case of both strongly interacting bosons and the impurity, an initially fast impurity exhibits slowly decaying oscillations in time before reaching a stationary state \cite{QFlutterNature,knap2014quantum}.
This phenomenon was dubbed quantum flutter. It was claimed that quantum flutter cannot be captured by a hydrodynamic theory and the Gross-Pitaevskii equation, and that it requires the strong coupling regime \cite{QFlutterNature}. Recently, we have shown that the description within the Gross-Pitaevskii equation reveals the velocity oscillations of a heavy fast impurity in a system of weakly interacting bosons \cite{Quench}. Moreover, the oscillations are much more pronounced for a weakly coupled impurity. 
	
In a system of weakly interacting bosons, contrary to the case of an attractively interacting impurity, the repulsive one exhibits a {damped} oscillations \cite{Quench}. Furthermore, these oscillations disappear at strong $\tilde{G}$, in contrast to the oscillating state that gets longer-lived as $|\tilde{G}|$ is increased. This behavior is shown in Fig.~\ref{fig:repulsive}.
The underlying mechanisms are also different. In the repulsive case, $G>0$, an effective attractive interaction between the  local depletion cloud and the impurity produces damped oscillations of the depletion cloud around the impurity until their positions coincide and they continue the motion together. For $G<0$, the density peak does not oscillate around the impurity, and its maximal density is situated at the impurity position during the impurity velocity oscillations.

\subsection{Ionic impurity}

The dynamics of a charged impurity interacting with bosons via an attractive, long-range potential was studied in Ref.~\cite{Ion1d}. For the impurity mass $M=m$ and the initial velocity $V_0=v$, it was found that a sufficiently strong impurity-boson interaction induces temporal impurity velocity oscillations before a stationary state is reached. However, in stark contrast to our findings, the oscillations reported in that study are heavily damped,  and disappear within the time window of  $10 \hbar/gn_0$. The underlying  mechanism identified in our work was not reported in their study, nor was the dependence of the oscillation properties -- and their domain of existence -- on the system parameters.  

It would be interesting to determine whether the dynamical trapping of the depletion cloud extends to ionic impurities, particularly in the heavy-impurity limit. 
The impact of the impurity mass and initial velocity on both the oscillatory and non-oscillatory dynamics remains to be studied for an ionic impurity. Moreover, the properties of the oscillations as a function of interaction strength have yet to be fully characterized.

\section{Conclusions \label{sec:disscusion}}
\begin{figure*}
\includegraphics[width=0.48\linewidth]{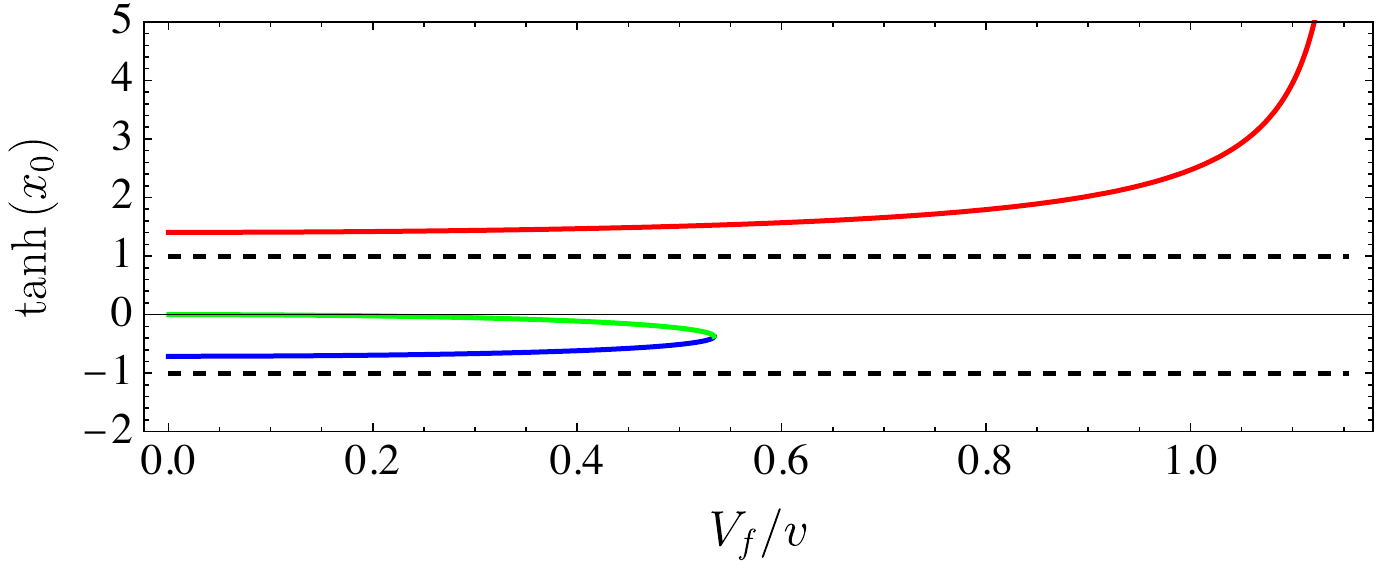} 
\includegraphics[width=0.48\linewidth]{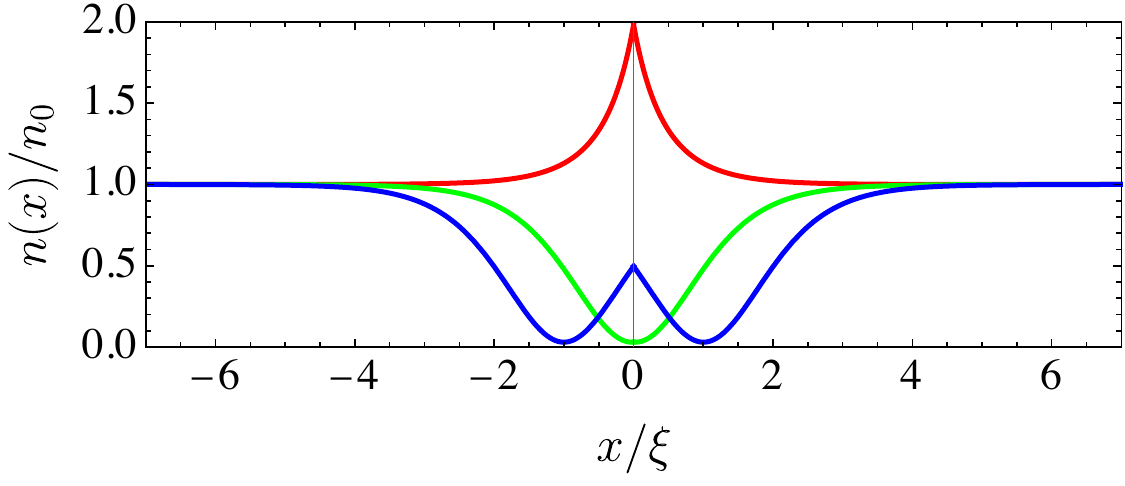} 
\caption{(left) Behavior of the solutions of \eq{eq:xo} as a function of the final impurity velocity. Here $M=3m$ and $\tilde{G}=-0.8$. (right) The corresponding density profile at $V_f=0.2 v$. The colour code is the same for both figures.}
\label{fig:xo}
\end{figure*}
In this work, we have studied the far-from-equilibrium dynamics of an attractively interacting impurity injected with a finite velocity into a ground state of a system of weakly-interacting homogeneous 1d bosons. 
We have analyzed the stationary states reached under two distinct protocols: (i) the adiabatically slow and (ii) the sudden switching on of the impurity-boson interaction. We have demonstrated that the state taking a soliton-like form—characterized by an increased boson density moving together with the impurity—possesses forbidden momentum sectors within protocol (i). The underlying cause stems from the requirement that the stationary impurity velocity must remain below the critical velocity and the absence of excitations within protocol (i), thereby imposing constraints on the total momentum. 
The critical velocity is independent on the impurity-boson coupling constant and equals the mass-scaled sound velocity. In protocol (ii), however, this state is reached at an arbitrary system momentum. This is facilitated by the transfer of momentum from the impurity to the bath via the emission of excitations that propagate away from the impurity. 

In Sec.~\ref{sec:relaxation}, we have characterized the relaxation dynamics of both the background bosons and the impurity under protocol (ii). We focused on the emitted excitations, the relevant timescales, and the final impurity velocity as a function of the initial velocity, the impurity mass, and the impurity-boson coupling strength.
We have uncovered a novel dynamical phenomenon resulting from a fast impurity with a mass close to or greater than the critical one, see Sec.~\ref{sec:resonant}. The impurity velocity undergoes undamped long-lived oscillations in time before reaching a stationary state. 
The underlying physical mechanism is the formation of a non-stationary state where the local boson depletion cloud gets temporarily localized and oscillates around the density peak situated at the impurity position. The interference between the depletion hole and the peak results in a force that provokes the impurity oscillations. The momenta of the impurity and the surrounding bosons oscillate in anti-phase, until the depletion cloud gets split into two parts situated on different sides of the impurity. Due to the repulsive impurity-depletion and the depletion-depletion interaction, the depletion holes liberate and move away in the opposite directions from the impurity, while locally a stationary state is reached. An increase of the initial impurity velocity, the impurity mass or the absolute value of the impurity-boson coupling is beneficial for the formation of the oscillating state. These parameters also determine the amplitude, the frequency and the lifetime of the oscillations,  see Sec.~\ref{sec:resonant}. Importantly, the stronger the impurity-boson attraction is, the longer the lifetime of the oscillating state is and the higher the frequency of the oscillations is.

Although the influence of quantum fluctuations is beyond the scope of this work, the phase coherence length  $\ell_{\phi}=\xi \exp{(2\sqrt{\pi^2/\gamma})}$ is huge for weakly interacting bosons \cite{Petrov}, and quantum fluctuations can be taken into account using the perturbative expansion (\ref{eq:expansion}) \cite{pitaevskii_bose-einstein_2003,sykes_drag_2009,CasimirNewJPhys}. While fluctuations are more prominent for light and weakly coupled impurities, we focused here on the complementary parameter region. Note that the short-distance attraction (repulsion) between a density peak and a depletion cloud  (depletion-depletion) is accurately characterized within the mean-field approach \cite{CasimirNewJPhys}. Thus, fluctuations are not expected to be relevant for the impurity oscillations. Fluctuations may lead to a weak damping of the oscillations, as in the case of the impurity Bloch oscillations \cite{Schecter_2016}. Cold atomic gases provide an ideal playground for the impurity physics \cite{2012quantum,Meinert945,JinPhysRevLett.117.055301,PhysRevLett.117.055302,HadzibabicPhysRevX.15.021070,grusdt2025impurities}, where the long-lived oscillations should be experimentally observable. 
Moreover, it would be valuable to verify whether the impurity dynamics discussed here extends to the regime of strong boson-boson interactions, as well as to higher-dimensional systems.

\section*{Acknowledgments}
	
This study has been partially supported through the EUR grant NanoX n° ANR-17-EURE-0009 in the framework of the ``Programme des Investissements d'Avenir".
	
\appendix

\section{Stationary solutions of \eq{eq:mean-field1}\label{sec:Appendix}}

Here, we consider all the stationary solutions of \eq{eq:mean-field1} that take the form (\ref{eq:MFsolution}). In the main text, we focused mostly on the properties of the physical ground-state solution that leads to an increased density at the impurity position, $n(0)>n_0$. However, note that all the expressions in terms of $x_0$ from Sec.~\ref{sec:Apolaron} apply also to these solutions.

There are three solutions of \eq{eq:xo}. The two of them have negative or vanishing $x_0$ and thus give $n(0)\leq n_0$. Their general behaviour is illustrated in Fig.~\ref{fig:xo}. Note that the two solutions with negative $x_0$ merge at the final impurity velocity $V_0<v_c$ given by
\begin{align}\label{eq:mergeing}
	\frac{-\tilde{G}}{\sqrt{1+m/M}}=\frac{\sqrt{1-20
			\tilde{V}_0^2-8 \tilde{V}_0^4+\left(1+8 \tilde{V}_0^2\right)^{3/2}}}{2 \sqrt{2}\tilde{V}_0}.
\end{align}
Here we have introduced $\tilde{V}_0=V_0/(v \sqrt{1+m/M})$.  
At higher values of $V_f$,  the solutions take complex values, and give unphysical complex values for the density. At $V_f=0$, the solutions read as $X=\tanh{x_0}$:
\begin{align}
 X_{1}&=0,\\
 X_{2/3}&=\frac{-\tilde{G}\pm \sqrt{4+4m/M+\tilde{G}^2}}{2\sqrt{1+m/M}}.
 \end{align}
Here, $X_1$, $X_2$ and $X_3$ correspond to the green, red and blue solutions, respectively.

The corresponding density profiles are shown in the right panel of Fig.~\ref{fig:xo}. Note that $\tanh{x_0}$ for the red curve diverges as $1/\sqrt{v_c-V_f}$ for $V_f\to v_c^-$, and the density at the impurity position is finite. The value of $x_0$ for the green curve is very small at $V_f=0.2 v$ and thus the peak at $x=0$ is not visible. As a result, the density profile looks like a soliton. Increasing $V_f$, the density at the impurity position of the green solution grows and of the blue one diminishes, until they merge. For the solution shown in blue, reducing $|\tilde{G}|$ leads to a further separation of the two depletion regions and the density at the impurity position grows.   Note that $n(0)$ reaches its maximal value $n_0$, when the depletion regions are infinitely far away as $\tilde{G}\to 0^-$. For the solution represented by the green colour, reducing $|\tilde{G}|$ leads to the opposite effect, the two depletion regions approach each others and the density at the impurity position diminishes. As a result the peak disappears and the density profile is given by a dark soliton.

\begin{figure}[h]
\includegraphics[width=0.49\linewidth]{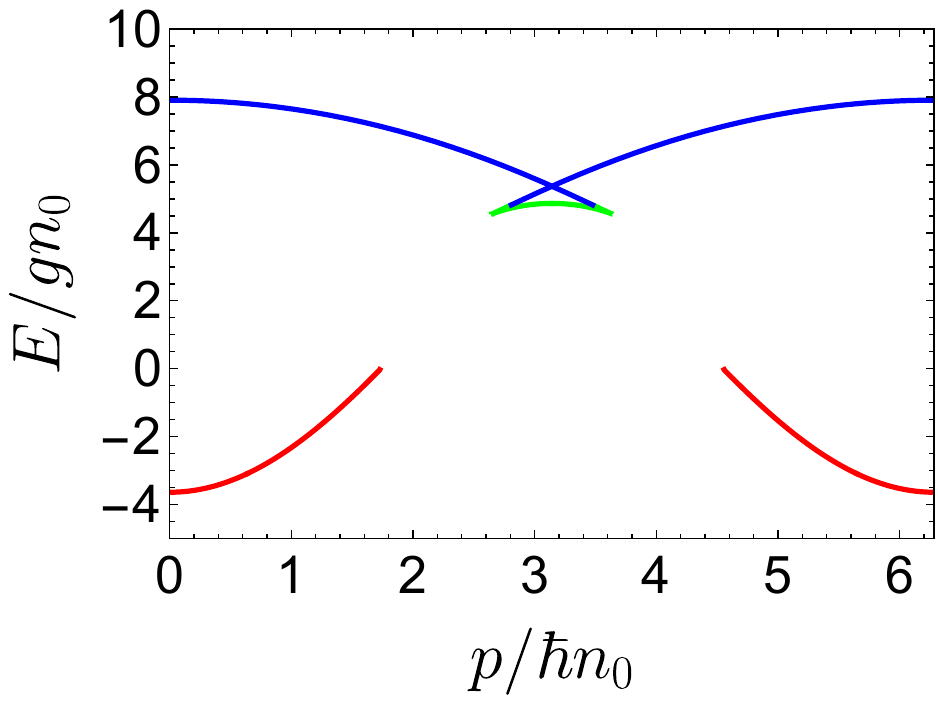} 
\includegraphics[width=0.49\linewidth]{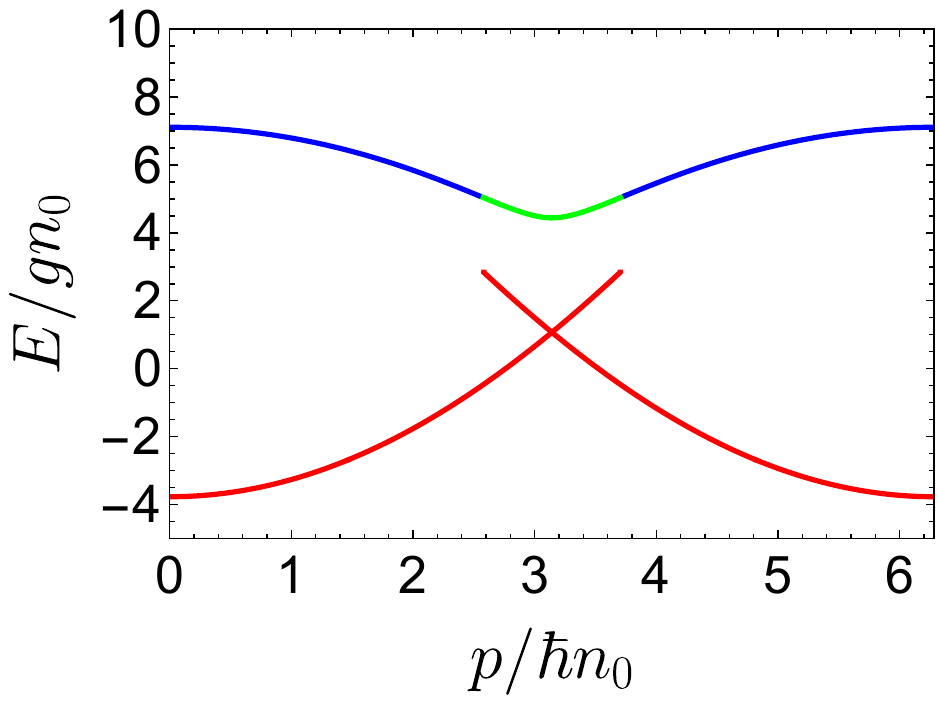}
\caption{Energy dispersion for (left) $M=3m$ and (right) $M=9m>M_c$. Here  $\tilde{G}=-0.8$ and $\gamma=0.1$. The colour code is the same as in Fig.~\ref{fig:xo}.}
\label{fig:energyxo}
\end{figure}

\begin{figure}[h]
\includegraphics[width=0.49\linewidth]{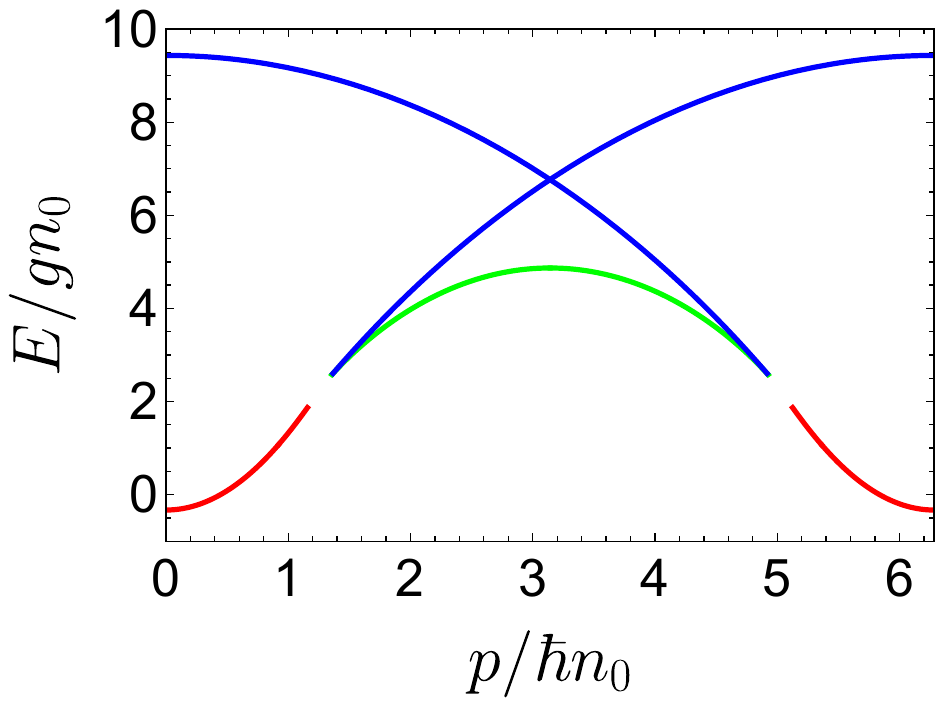} 
\includegraphics[width=0.49\linewidth]{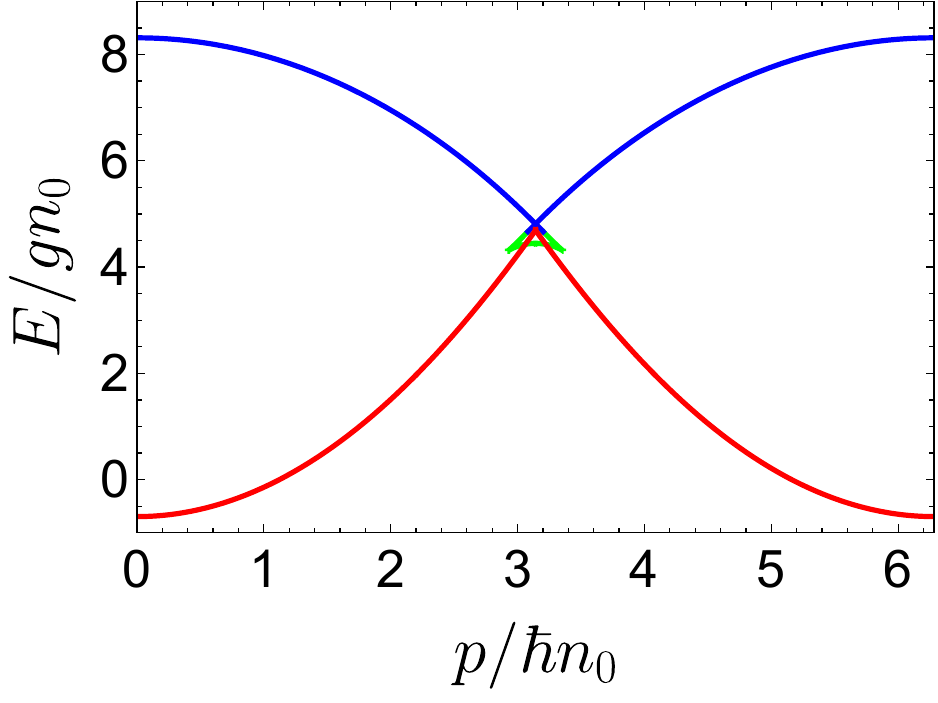}
\caption{Energy dispersion for (left) $M=3m$, $\tilde{G}=-0.1$ and $\gamma=0.1$, and (right) $M=9m>M_c$, $\tilde{G}=-0.2$ and $\gamma=0.1$. The colour code is the same as in Fig.~\ref{fig:xo}.}
\label{fig:special}
\end{figure}

The energy dispersion of the stationary solutions is shown in Fig.~\ref{fig:energyxo} for impurity masses lower and higher than the critical one. In a general case, we note that the energy branches do not merge, as shown in Fig.~\ref{fig:energyxo}. This is a typical energy dispersion, with the physical solution shown in the red colour having the lowest energy at all momenta. 
Reducing $|\tilde{G}|$, the energy branches approach each other and finally merge. However, only the red one tends to $p^2/(2M)$ in the limit of vanishingly small coupling. This behaviour is illustrated in the left panel of Fig.~\ref{fig:special}. The influence on the case $M>M_c$ is shown in the right panel of Fig.~\ref{fig:special}. There, the energies of solutions represented by  the blue and the green  colour can be lower than the red one for momenta around $p=\pi \hbar n_0$.

The two types of stationary density profiles, one with a $n(0)>n_0$ and another with $n(0)<n_0$ have been reported for an infinitely heavy impurity \cite{PhysRevA.64.033602}, although their explicit form beyond the small-$G$ expansion was not provided. Note that we find that the solution with $n(0)>n_0$ exists at all velocities $|V_f|<v_c$, while the author reports only the interval $v_c>|V_f|>V_0$.  Moreover, at $M=\infty$, we expect that the solution shown in green colour is unstable, since in the limit $\tilde{G}\to 0^-$ it does not give $n(x)=n_0$. The above described behaviour of the blue solution as $\tilde{G}$ diminishes suggests that it is also physically unexpected.


%

\end{document}